\documentclass[11pt,leqno]{article}
\usepackage{latexsym}
\usepackage{amsfonts}
\usepackage{amsmath}
\usepackage{amssymb}

\newlength{\filength}
\newsavebox{\gcbox}
\sbox{\gcbox}{\framebox[\filength]{\rule{0ex}{2ex}}}

\newtheorem{theorem}{Theorem}[section]
\newtheorem{corollary}[theorem]{Corollary}

\newcommand{\qedblob}{\mbox{\rule[-1.5pt]{5pt}{10.5pt}}}
\def\literalqed{{\ \nolinebreak\hfill\mbox{\qedblob\quad}}}

\def\qed{\literalqed}

\newtheorem{proposition}[theorem]{Proposition}
\newtheorem{definition}[theorem]{Definition}
\makeatletter
\def\@citex[#1]#2{\if@filesw\immediate\write\@auxout{\string\citation{#2}}\fi
  \def\@citea{}\@cite{\@for\@citeb:=#2\do
    {\@citea\def\@citea{,\linebreak[0]}\@ifundefined
       {b@\@citeb}{{\bf ?}\@warning
       {Citation `\@citeb' on page \thepage \space undefined}}%
\hbox{\csname b@\@citeb\endcsname}}}{#1}}
\makeatother

\newcommand{\naturalnumber}{\ensuremath{{  \mathbb{N} }}}



\newcommand{\p}{{\rm P}}

\newcommand{\np}{{\rm NP}}

\newcommand{\pair}[1]{\mathopen\langle{#1}\mathclose\rangle}

\newcommand{\sigmastar}{\ensuremath{\Sigma^\ast}}

\def\land{{\; \wedge \;}}
\def\lor{{\; \vee \;}}

\newcommand{\fp}{{\rm FP}}
\def\nats{\naturalnumber}

\newcommand\Lora{\, \Longrightarrow \ }

\newcommand{\image}{\mbox{\rm{}image}}
\newcommand{\domain}{\mbox{\rm{}domain}}

\title{If $\p \neq \np$ then Some Strongly Noninvertible Functions are
  Invertible}

\author{
Lane A. Hemaspaandra\,\thanks{Department of Computer Science,
University of Rochester,
Rochester, NY 14627, USA.
Email: {\tt lane@cs.rochester.edu}.
Supported in part 
by grants NSF-CCR-9322513 and 
NSF-INT-9815095/DAAD-315-PPP-g\"{u}-ab.
Work done in part while visiting
the University of W\"urzburg.}
\\
University of Rochester
\and 
Kari Pasanen\,\thanks{Nokia Networks,
P.O.\ Box 12,
FIN-40101 Jyv\"{a}skyl\"{a},
Finland.
Email: {\tt kari.pasanen@nokia.com}.  
}
\\
Nokia Networks and University of Jyv\"{a}skyl\"{a}
\and 
J\"{o}rg Rothe\,\thanks{Abteilung f\"ur Informatik,
Heinrich-Heine-Universit\"at D\"usseldorf,
40225 D\"usseldorf, Germany.
Email: {\tt rothe@informatik.uni-jena.de}.
Supported in part by grant
NSF-INT-9815095/DAAD-315-PPP-g\"{u}-ab and a Hei\-sen\-berg Fellowship
of the Deut\-sche For\-schungs\-gemein\-schaft.}
\\
Heinrich-Heine-Universit\"at D\"usseldorf
}

\date{October 6, 2000}

\begin{document}

\sloppy

\bibliographystyle{alpha}

\maketitle

\begin{abstract}
  Rabi, Rivest, and Sherman alter the
  standard notion of noninvertibility to a new notion they call
  strong noninvertibility, and show---via explicit cryptographic
  protocols for secret-key agreement 
  (\cite{rab-she:t-no-URL:aowf,rab-she:j:aowf} attribute this
  to Rivest and Sherman) and digital 
  signatures~\cite{rab-she:t-no-URL:aowf,rab-she:j:aowf}---that
  strongly noninvertible functions would be very useful components
  in protocol design.  Their definition of strong noninvertibility has
  a small twist (``respecting the argument given'') that is needed to
  ensure cryptographic usefulness.  In this paper, we show that this
  small twist has a large, unexpected consequence: Unless $\p = \np$,
  some strongly noninvertible functions are invertible. 

\smallskip\noindent
{\bf Topic:} Computational and Structural Complexity.
\end{abstract}

\section{Introduction}

Rabi, Rivest, and Sherman developed novel cryptographic protocols that
require one-way functions with algebraic properties such as
associativity (see~\cite{rab-she:t-no-URL:aowf,rab-she:j:aowf} and 
the attributions and references therein, 
esp.~\cite{she:phd,kal-riv-she:j:is-des-a-group}).
Motivated by these protocols, they initiated the study of two-argument
(2-ary, for short) one-way functions in worst-case cryptography.  
To preclude certain types of attacks, their protocols require
one-way functions that are not invertible in polynomial time even when
the adversary is given not just the function's output but also
one of the function's inputs.  Calling this property of one-way functions
``strong noninvertibility'' (or ``strongness,'' for short), they left
as an open problem whether there is any evidence---e.g., any plausible
complexity-theoretic hypothesis---ensuring the existence of one-way
functions with all the properties the protocols require, namely 
ensuring the existence of total,
commutative, associative one-way functions that are strongly
noninvertible.  This problem was recently solved by Hemaspaandra and
Rothe~\cite{hem-rot:j:aowf} who show that if $\p \neq \np$ then such
one-way functions do exist.

Unfortunately, Hemaspaandra and
  Rothe~\cite{hem-rot:j:aowf} write: ``Rabi and
  Sherman~\cite{rab-she:j:aowf} also introduce the notion of {\em
    strong\/} one-way functions---2-ary one-way functions that are
  hard to invert even if one of their arguments is given.  Strongness
  implies one-way-ness.''~~The latter sentence could be very 
  generously read as meaning ``strong,
  one-way functions'' when it speaks of ``strongness,'' especially since
  strongness alone, by definition, does not even require honesty, and
  without honesty the sentence quoted above would be provably, trivially,
  false.  
  However, a more natural reading is that~\cite{hem-rot:j:aowf} is 
  assuming that strongly noninvertible
  functions are always noninvertible.  The main result of
  the present paper is that if $\p \neq \np$ then this is
  untrue.  So, even when one has proven a function to be strongly
  noninvertible, one must not merely claim that noninvertibility
  automatically holds (as it may not), but rather one must prove the
  noninvertibility.\footnote{Since in~\cite{hem-rot:j:aowf} only strong 
noninvertibility is
explicitly proven, one might worry that the functions constructed in its
proofs may be invertible.
Fortunately, the constructions in the proofs
in~\cite{hem-rot:j:aowf} do easily support and implicitly give
noninvertibility as well;
thus, all the claims of~\cite{hem-rot:j:aowf} remain correct.
Most crucially, on page~654
  of~\cite{hem-rot:j:aowf}, inverting the output $\pair{x,x}$ in 
  polynomial time would
  give strings containing one witness for membership of $x$ in the
  given set in $\np - \p$ (if there are any such witnesses), which is
  impossible.
}

In the present paper, we study appropriately honest, 
polynomial-time computable 2-ary functions.  We prove that
if $\p \neq \np$ then there exist strongly
noninvertible such functions that are invertible (see 
Section~\ref{s:defs} for precise definitions).  
This is a rather surprising result that at first might seem paradoxical.
To paint a full picture of what happens if $\p \neq \np$,
we also show the (nonsurprising) result 
that if $\p \neq \np$ then there exist 
appropriately honest, 
polynomial-time computable 2-ary functions that are noninvertible, yet not
strongly noninvertible. 

So, why is the surprising, paradoxical-seeming result (that
if $\p \neq \np$ then some strongly
noninvertible functions are invertible) even possible?
Let us informally explain.
Let $\sigma$ be a 2-ary function.  We say $\sigma$ is noninvertible if
there is no polynomial-time inverter that, given an image element
$z$ of~$\sigma$, outputs some preimage of~$z$.  We say $\sigma$ is
strongly noninvertible if even when, in addition to any image element
$z$ of~$\sigma$, one argument of $\sigma$ is given such that there exists
another string with which this argument 
is mapped to~$z$, computing one such
other argument is not a polynomial-time task.  So, why does strongness
alone not outright imply noninvertibility?  One might be tempted to
think that from some given polynomial-time inverter $g$ witnessing the
invertibility of $\sigma$ one could construct
polynomial-time inverters $g_1$ and $g_2$ such that $g_i$ inverts
$\sigma$ in polynomial time even when the $i$th argument is given
(see Definition~\ref{d:strong-oneway} for the formal details).
This approach does not work.  In
particular, it is not clear how to define~$g_1$ when given an output
$z$ of $\sigma$ and a first argument $a$ that together with a
corresponding second argument is mapped to~$z$, yet $a$ is not the
first component of~$g(z)$.  (In fact, our main theorem implies that
{\em no\/} approach can in general
accomplish the desired 
transformation from $g$ to~$g_1$, unless $\p = \np$.)

But then, why don't we use a different notion of strongness that
automatically implies noninvertibility?  
The answer is that the definitional subtlety that opens the door to
the unexpected behavior is absolutely essential to the cryptographic
protocols for which Rabi, Rivest, and Sherman created the notion 
in the first place.
For example, suppose one were
tempted to redefine ``strongly noninvertible'' with the following quite
different notion: $\sigma$ is ``strongly noninvertible'' if, given any
image element $z$ of $\sigma$ and any one argument of $\sigma$ such that
there exists another string with which this argument is mapped to~$z$,
computing {\em any preimage of $z$\/} (as opposed to ``any other
argument respecting the argument given'') is not a polynomial-time
task.  
The problem with this redefinition is that it completely loses
the core of why strongness precludes direct attacks
against the protocols of Rabi, Rivest, and
Sherman.  We will call the
just-defined notion ``overstrongness,'' as it seems to be
overrestrictive in terms of motivation---and we will prove that if
$\p \neq \np$ then overstrongness indeed is a properly more restrictive
notion than strongness.

\section{Definitions}
\label{s:defs}

Fix the binary alphabet $\Sigma = \{0,1\}$. Let $\epsilon$ denote 
the empty string.  Let $\pair{\cdot, \cdot}
: \sigmastar \times \sigmastar \rightarrow\, \sigmastar$ be some
standard pairing function, that is, some total, polynomial-time
computable bijection that has polynomial-time computable inverses and
is nondecreasing in each argument when the other argument is fixed.
Let FP denote the set of all polynomial-time computable total
functions.  The standard definition of one-way-ness used here
is essentially 
due to Grollmann and Selman~\cite{gro-sel:j:complexity-measures}
(except that they require one-way functions to be one-to-one); as in
the papers~\cite{rab-she:j:aowf,hem-rot:j:aowf,hom:t:low-ambiguity-aowf}, 
their notion is tailored below to the case of 2-ary functions.

\begin{definition}
{\rm{}\cite{gro-sel:j:complexity-measures,rab-she:j:aowf,hem-rot:j:aowf}}
\quad
\label{d:oneway}
  Let $\rho : \sigmastar \times \sigmastar \rightarrow \sigmastar$
  be any (possibly nontotal, possibly many-to-one) 2-ary function.
\begin{enumerate}
\item 
  We say $\rho$ is {\em honest\/} if and only if there exists a
  polynomial $q$ such that:
\[
(\forall z \in \image(\rho))\, (\exists (a,b) \in \domain(\rho))\,
[|a| + |b| \leq q(|z|) \land \rho(a,b) = z].
\]

\item 
We say $\rho$ is {\em (polynomial-time) noninvertible\/} if and
  only if the following does not hold:
\[
(\exists g \in \fp)\, (\forall z \in \image(\rho))\, [\rho(g(z)) = z].
\]

\item We say $\rho$ is {\em one-way\/} if and only if it is honest,
  polynomial-time computable, and noninvertible.
\end{enumerate}
\end{definition}

We now define strong noninvertibility (or strongness), which is a stand-alone
property (i.e., with one-way-ness not necessarily required) of
2-ary functions.  If one wants to discuss strongness in a nontrivial
way, one needs some type of honesty that is suitable for strongness.
To this end, we introduce below, in addition to honesty as defined
above, the notion of s-honesty.\footnote{The strongly noninvertible 
functions in~\cite{hem-rot:j:aowf} 
clearly are all s-honest, notwithstanding that s-honesty is not
explicitly discussed in~\cite{hem-rot:j:aowf} 
(or~\cite{rab-she:j:aowf,rab-she:t-no-URL:aowf}).
}

\begin{definition}
{\rm{}(see, essentially,
\cite{rab-she:j:aowf,hem-rot:j:aowf})}
\quad
\label{d:strong-oneway}
  Let $\sigma : \sigmastar \times \sigmastar \rightarrow \sigmastar$
  be any (possibly nontotal, possibly many-to-one) 2-ary function.
\begin{enumerate}
\item 
  We say $\sigma$ is {\em s-honest\/} if and only if there exists a
  polynomial $q$ such that both~{\rm{}(a)} and~{\rm{}(b)} hold:
\begin{description}
\item[(a)]
$(\forall z, a : (\exists b)\, [\sigma(a,b) = z])\, (\exists b')\,
[|b'| \leq q(|z| + |a|) \land \sigma(a,b') = z]$.
\item[(b)]
$(\forall z, b : (\exists a)\, [\sigma(a,b) = z])\, (\exists a')\,
[|a'| \leq q(|z| + |b|) \land \sigma(a',b) = z]$.
\end{description}

\item 
We say $\sigma$ is {\em (polynomial-time) invertible
    with respect to the first argument\/}  if and only if 
\begin{eqnarray*}
 & & (\exists g_1 \in \fp)\, (\forall z \in \image(\sigma))\, (\forall a, b
  : (a,b) \in \domain(\sigma) \land \sigma(a,b) = z) \\
 & & [\sigma(a, g_1(\pair{a,z})) = z].
\end{eqnarray*}

\item 
We say $\sigma$ is {\em (polynomial-time) invertible
    with respect to the second argument\/}  if and only if 
\begin{eqnarray*}
 & & (\exists g_2 \in \fp)\, (\forall z\in \image(\sigma))\, (\forall a, b
  : (a,b) \in \domain(\sigma) \land \sigma(a,b) = z) \\
 & &   [\sigma(g_2(\pair{b,z}), b) = z].
\end{eqnarray*}

\item We say $\sigma$ is {\em strongly
    noninvertible\/} if and only if $\sigma$ is neither invertible
    with respect to the first argument nor invertible with respect to
    the second argument.

\item We say $\sigma$ is {\em strongly one-way\/} if and only if it is
  s-honest, polynomial-time computable, and strongly noninvertible.
\end{enumerate}
\end{definition}

It is easy to see that there are honest, polynomial-time computable
2-ary functions that are not s-honest,\footnote{For example, consider
the function $\rho : \sigmastar \times \sigmastar \rightarrow
\sigmastar$ defined by $\rho(a,b) = 1^{\lceil \log \log (\max(|b|,2))\rceil}$
if $a = 0$, and $\rho(a,b) = ab$ if $a \neq 0$.  This function is 
honest (as proven by $\rho(\epsilon, x) = x$) but
is not s-honest, since for any given polynomial $q$ there are
strings $b \in \sigmastar$ and $z = 1^{\lceil \log \log (\max(|b|,2))\rceil}$
with $\rho(0,b) = z$, but the smallest $b' \in \sigmastar$ with
$\rho(0,b') = z$ satisfies $|b'| > q(|z| + |0|) = q(\lceil \log \log
(\max(|b|,2))\rceil + 1)$.  
} 
and that there are s-honest, polynomial-time computable 2-ary
functions that are not honest.\footnote{For example, consider the
function $\sigma : \sigmastar \times \sigmastar \rightarrow
\sigmastar$ that is defined by $\sigma(a,b) = 1^{\lceil \log \log
(\max(|a|,2))\rceil}$ if $|a| = |b|$, and that is undefined otherwise.  This
function is s-honest but not honest.
} 

For completeness, we also give a formal definition of
the notion of overstrongness
mentioned in the last paragraph of the introduction.
Note that overstrongness implies both noninvertibility and
strong noninvertibility.

\begin{definition}
\label{d:c-strong}
Let $\sigma : \sigmastar \times \sigmastar \rightarrow \sigmastar$ be
any (possibly nontotal, possibly many-to-one) 2-ary function.
We say $\sigma$ is {\em overstrong\/} if and only if for no $f \in
\fp$ with $f : \{1,2\} \times \sigmastar \times \sigmastar \rightarrow
\sigmastar \times \sigmastar$ does it hold that for each $i \in \{1,2\}$
and for all strings $z, a \in \sigmastar$:
\[
((\exists b \in \sigmastar) [(\sigma(a,b) = z \land i = 1) \lor
(\sigma(b,a) = z \land i = 2)]) \Lora \sigma(f(i,z,a)) = z .
\]
\end{definition}

\section{On Inverting Strongly Noninvertible Functions}

It is well-known (see, e.g., \cite{sel:j:one-way,bal-dia-gab:b:sctI:95}) 
that 1-ary
one-way functions exist if and only if $\p \neq \np$; as mentioned
in~\cite{hem-rot:j:aowf,rab-she:j:aowf}, 
the standard method to prove this result can
also be used to prove the analogous result for 2-ary one-way
functions.

\begin{theorem}
\label{thm:a}
{\rm{}(see~\cite{hem-rot:j:aowf,rab-she:j:aowf})}
$\p \neq \np$ 
if and only if total 2-ary one-way functions exist.
\end{theorem}

Now we show the main, and most surprising, result of this paper: 
If $\p \neq \np$
then one can invert some functions that are strongly noninvertible.

\begin{theorem}
\label{thm:c}
If $\p \neq \np$ then there exists a total, honest 2-ary function that is
a strongly one-way function but not a one-way function.
\end{theorem}

\noindent
{\bf Proof.}\quad Assuming $\p \neq \np$, 
by Theorem~\ref{thm:a} there exists
a total 2-ary one-way function $\rho$.
Define a
function $\sigma : \sigmastar \times \sigmastar \rightarrow
\sigmastar$ as follows:
\[
\sigma(a,b) = \left\{
\begin{array}{ll}
0\rho(x,y) & \mbox{if $(\exists x,y,z \in \sigmastar)\, [
a = 1\pair{x,y} \land b = 0z]$} \\ 
0\rho(y,z) & \mbox{if $(\exists x,y,z \in \sigmastar)\, [
a = 0x \land b = 1\pair{y,z}]$} \\ 
1xy & \mbox{if $(\exists x,y \in \sigmastar)\, 
            [(a=0x \land b=0y) \lor (a=1x \land b=1y)]$} \\ 
ab & \mbox{if $a=\epsilon \lor b=\epsilon$.}
\end{array}
\right.
\]

It is a matter of routine to check 
that $\sigma$ is polynomial-time computable, total, honest, and s-honest
(regardless of whether or not~$\rho$, which is honest, is s-honest).

If one could invert $\sigma$ with respect to one of its arguments 
then one could
invert~$\rho$, contradicting that $\rho$ is a one-way function.  In
particular, supposing $\sigma$ is invertible with respect to the first
argument via inverter $g_1 \in \fp$, we can use $g_1$ to define a function $g
\in \fp$ that inverts~$\rho$.  To see this, note that given any $w \in
\image(\rho)$ with $w \neq \epsilon$, $g_1(\pair{0,0w})$ must yield a
string of the form $b = 1\pair{y,z}$ with $\rho(y,z) = w$.  Thus,
$\sigma$ is not invertible with respect to the first argument.  An
analogous argument shows that $\sigma$ is not invertible with respect
to the second argument.  Thus, $\sigma$ is strongly noninvertible.
However, $\sigma$ is invertible, since every string $z \in
\image(\sigma)$ has an inverse of the form $(\epsilon,z)$; so, the FP
function mapping any given string $z$ to $(\epsilon,z)$ is an inverter
for~$\sigma$.  Hence, $\sigma$ is not a one-way function.~\qed

\medskip

The converse of Theorem~\ref{thm:c} immediately holds,
as do the converses of Proposition~\ref{p:new}, Corollary~\ref{cor:e}, 
and Theorems~\ref{thm:double-d}, \ref{thm:d}, and~\ref{thm:c-strong}.
However, although all these results in fact are equivalences, 
we will focus on only the interesting implication direction.

For completeness, we mention in passing that, assuming $\p \neq \np$, one
can construct functions that---unlike the function constructed in the
proof of Theorem~\ref{thm:c}---are strongly one-way {\em and\/}
one-way.  An example of such a function is the following modification
$\hat{\sigma}$ of the function $\sigma$ constructed in the proof of
Theorem~\ref{thm:c}.  As in that proof, let $\rho$ be a total 2-ary one-way
function, and define function $\hat{\sigma} : \sigmastar \times
\sigmastar \rightarrow \sigmastar$ by
\[
\hat{\sigma}(a,b) = \left\{
\begin{array}{ll}
0\rho(x,y) & \mbox{if $(\exists x,y,z \in \sigmastar)\, [
a = 1\pair{x,y} \land b = 0z]$} \\ 
0\rho(y,z) & \mbox{if $(\exists x,y,z \in \sigmastar)\, [
a = 0x \land b = 1\pair{y,z}]$} \\ 
1ab & \mbox{otherwise.} 
\end{array}
\right.
\]
Note that $\hat{\sigma}$ even is overstrong; hence, $\hat{\sigma}$ is
both noninvertible and strongly noninvertible.  That is:

\begin{proposition}
\label{p:new}
If $\p \neq \np$ then there exists a total, honest, s-honest, 2-ary 
overstrong function.  (It follows that 
if $\p \neq \np$ then there exists a total 2-ary function that is
one-way and strongly one-way.)
\end{proposition}

Corollary~\ref{cor:e} below shows that if $\p \neq \np$ then there is
an s-honest 2-ary one-way function that is not strongly one-way.  
First, we establish a result that is slightly stronger:
For a function to be not strongly noninvertible, it is enough that
it is invertible with respect to at least one of its arguments.
The function $\sigma$ to be constructed in the
proof of Theorem~\ref{thm:double-d} below
even is invertible with respect to each of its arguments.

\begin{theorem}
\label{thm:double-d}
If $\p \neq \np$ then there exists a total,
s-honest 2-ary
one-way function $\sigma$ such that $\sigma$
is invertible with respect to its first argument and $\sigma$
is invertible with respect to its second argument.
\end{theorem}

\noindent
{\bf Proof.}\quad It is well-known (\cite[Prop.~1]{sel:j:one-way},
in light of the many-to-one analog of his comment 
\cite[p.~209]{sel:j:one-way} about totality) 
that under the assumption $\p \neq \np$ there exists
a total 1-ary
one-way function $\rho : \sigmastar \rightarrow \sigmastar$.  Define a
function $\sigma : \sigmastar \times \sigmastar \rightarrow
\sigmastar$ as follows:
\[
\sigma(a,b) = \left\{
\begin{array}{ll}
1\rho(a) & \mbox{if $a = b$} \\ 
0ab & \mbox{if $a \neq b$.}
\end{array}
\right.
\]
Note that $\sigma$ is polynomial-time computable, total,
s-honest,
and 
honest.  If $\sigma$ were invertible in polynomial time
then $\rho$ would be too; so, $\sigma$ is a one-way function.  However,
$\sigma$ is invertible with respect to each
of its arguments.  For an inverter with respect to the first argument,
consider
the function $g_1 : \sigmastar \rightarrow \sigmastar$ defined by
\[
g_1(x) = \left\{
\begin{array}{ll}
b & \mbox{if $(\exists a, b, z \in \sigmastar)\, 
         [x = \pair{a,0z} \land z = ab]$} \\ 
a & \mbox{if $(\exists a, z \in \sigmastar)\, 
         [x = \pair{a,1z}]$} \\
\epsilon & \mbox{otherwise.}
\end{array}
\right.
\]
Clearly, $g_1 \in \fp$.
Note that for every $y \in
\image(\sigma)$ and for every $a \in \sigmastar$ for which there
exists some $b \in \sigmastar$ with $\sigma(a,b) = y$, it holds that
$\sigma(a, g_1(\pair{a,y})) = y$, completing the proof that
$\sigma$ is invertible with respect to the first argument. 
To see that $\sigma$ also is invertible with respect to the 
second argument, an analogous construction (with the
roles of the first and the second argument interchanged) works to give
an inverter $g_2$ for a fixed second argument.~\qed

\begin{corollary}
\label{cor:e}
If $\p \neq \np$ then there exists a total, s-honest 2-ary one-way function
that is not strongly one-way.
\end{corollary}

One might wonder whether functions that are not strongly noninvertible
(which means they are invertible with respect to at least one of their
arguments) outright must be invertible with respect to both of their
arguments.  The following result states that this is not the case,
unless $\p = \np$.  

\begin{theorem}
\label{thm:d}
If $\p \neq \np$ then there exists a total, s-honest 2-ary one-way
function that is invertible with respect to one of its arguments
(thus, it is not strongly one-way), yet that is not invertible with
respect to its other argument.
\end{theorem}

\noindent
{\bf Proof.}\quad Assuming $\p \neq \np$, 
by Theorem~\ref{thm:a} there exists
a total 2-ary one-way function, call it~$\rho$.  Since our 
pairing function is onto and one-to-one, and its 
inverses are efficiently computable, the functions---$\pi_1$ and 
$\pi_2$---mapping from 
each string in
$\sigmastar$ to that string's first and second components when interpreted
as a pair are well-defined, total, polynomial-time functions;  for all 
$b \in\sigmastar$, 
$b = \pair{\pi_1(b), \pi_2(b)}$.
Define a function $\sigma :
\sigmastar \times \sigmastar \rightarrow \sigmastar$ as follows:
\[
\sigma(a,b) = 
\rho(\pi_1(b), \pi_2(b))
\]
It is clear that $\sigma$ is honest (via $\rho$'s honesty) and s-honest.
Let $a_0$ be any fixed string, and define $g_2(w) = a_0$ for all
strings~$w$.  Clearly, $g_2 \in \fp$.  The definition of $\sigma$
implies that for each $z = \rho(x,y) \in \image(\sigma)$ and for each
$b \in \sigmastar$ such that $\sigma(a,b) = z$ for some $a \in
\sigmastar$, it also holds that $\sigma(a_0,b) = z$. Thus, $\sigma$
is invertible with respect to the second argument via~$g_2$.
However, if $\sigma$ were also invertible with respect to the first argument
via some function $g_1 \in \fp$, then $g_1$ could be used to
invert~$\rho$, which would contradict the noninvertibility of~$\rho$.
Hence, $\sigma$ is invertible with respect to its first, yet not with 
respect to its second argument.  Analogously, we can define a
function that is invertible with respect to its second argument, 
yet not with respect to its first argument.~\qed

\medskip

Finally, let us turn to the notion of overstrongness (see
Definition~\ref{d:c-strong}) mentioned in the last paragraph of the
introduction.  As noted there, this notion is not less restrictive
than either noninvertibility or strong noninvertibility, and so
if a given polynomial-time computable, honest, s-honest function is
overstrong then it certainly is both one-way and strongly one-way.
Notwithstanding the fact that---as we have argued---overstrongness is
not well-motivated by the cryptographic protocols of Rabi, Rivest, and
Sherman~\cite{rab-she:j:aowf}, for the purpose of showing that the
notions do not collapse, we will prove that the converse does not hold,
unless $\p = \np$.

\begin{theorem}
\label{thm:c-strong}
If $\p \neq \np$ then there exists a total, honest, s-honest
2-ary function that is noninvertible and strongly noninvertible
but that is not overstrong.
\end{theorem}

\noindent
{\bf Proof.}\quad Assume $\p \neq \np$.  
It is known (see~\cite{sel:j:one-way}) that this assumption implies that 
total 1-ary one-way functions exist.
Let $\widehat{\rho}$ be one such function, and
let $\widehat{\rho}$ be such that it additionally satisfies
$(\exists r \geq 2)\, (\forall x \in \sigmastar)\, 
[|\widehat{\rho}(x)| = |x|^r + r]$.  Henceforth, $r$ will denote this
value~$r$.  That this condition can be required follows easily from the
standard ``accepting-paths-based'' proofs that $\p \neq \np$ implies
the existence of total 1-ary one-way functions. 

Define a total function
$\rho : \sigmastar \rightarrow \sigmastar$ as follows:
\[
\rho(a) = \left\{
\begin{array}{ll}
1\widehat{\rho}(x) & \mbox{if $(\exists x \in \sigmastar)\, [a = 1x]$} \\ 
a & \mbox{if $(\exists x \in \sigmastar)\, [a = 0x]$} \\ 
\epsilon & \mbox{if $a = \epsilon$.} 
\end{array}
\right.
\]
Note that $\rho$ is a 1-ary, total one-way function satisfying that
for each $i \geq 0$, $\rho(0^i) = 0^i$.  Now define the total function
$\sigma : \sigmastar \times \sigmastar \rightarrow \sigmastar$ as
follows:
\[
\sigma(a,b) = \left\{
\begin{array}{ll}
1\pair{\rho(x),0^{|y|}} & \mbox{if $(\exists x,y \in \sigmastar)\, 
                  [|x| = |y| \land a = 0\pair{x,y} = b]$} \\ 
1\pair{\rho(x),0^{|y|}} & \mbox{if $(\exists x,y \in \sigmastar)\, 
                  [|x| = |y| \land a = 1\pair{x,0y} 
                   \land b = 1\pair{x,1\widehat{\rho}(y)}]$} \\ 
1\pair{\rho(x),0^{|y|}} & \mbox{if $(\exists x,y \in \sigmastar)\, 
                  [|x| = |y| \land a = 1\pair{x,1\widehat{\rho}(y)}
                   \land b = 1\pair{x,0y}]$} \\ 
0\pair{a,b} & \mbox{otherwise.}
\end{array}
\right.
\]
Clearly, $\sigma$ is polynomial-time computable, honest, s-honest, and
commutative.  If $\sigma$ were invertible, $\rho$ would be too.  Thus,
$\sigma$ is a one-way function.

Note that $\sigma$ is strongly noninvertible, for if it could
be inverted with respect to either argument then $\widehat{\rho}$
could be inverted too.  Suppose, for example, $\sigma$ were invertible with
respect to the first argument via inverter $g_1 \in \fp$.  Then
$\widehat{\rho}$ could be inverted as follows.  Given any $z \in
\sigmastar$, if there is no $k \in \nats$ with $k^r + r = |z|$,
there is no inverse of $z$ under~$\widehat{\rho}$; so, in that case we
may output anything.  Otherwise (i.e., there is a $k \in \nats$
with $k^r + r = |z|$), run $g_1$ on input $\pair{a, w}$, where
$a = 1\pair{0^k, 1z}$ and $w = 1\pair{0^{k}, 0^{k}}$.  By the definition
of~$\sigma$, if $z \in \image(\widehat{\rho})$, the result of
$g_1(\pair{a, w})$ must be of the form $1\pair{0^k, 0\hat{z}}$ for
some preimage $\hat{z}$ of $z$ under~$\widehat{\rho}$, and we can
verify this by running $\widehat{\rho}$ on input $\hat{z}$ and
checking whether or not $\widehat{\rho}(\hat{z}) = z$.  A similar
argument shows that $\sigma$ is not invertible with respect to the
second argument.  Hence, $\sigma$ is strongly one-way.

Finally, we claim that $\sigma$ is not overstrong.  Here is what an
inverter $f$ does when given $i = 1$,\footnote{Since $\sigma$ is
commutative, this implicitly also shows how to handle the case $i =
2$.
} 
an alleged first argument $a \in \sigmastar$ of~$\sigma$, and an alleged 
output~$z \in \sigmastar$ of~$\sigma$:
\[
f(1,a,z) = \left\{
\begin{array}{ll}
(x,y) & \mbox{if $(\exists x,y \in \sigmastar)\, 
                  [z = 0\pair{x,y}]$} \\ 
(a,a) & \mbox{if $(\exists x,y \in \sigmastar)\, (\exists m \in \nats)\, 
                  [a = 0x \land z = 1\pair{y,0^m}]$} \\ 
(0\pair{w,w},0\pair{w,w}) & 
          \mbox{if $(\exists w,x,y \in \sigmastar)\, (\exists m \in \nats)\, 
                  [a = 1\pair{w, 0x} \land z = 1\pair{y,0^m}]$} \\ 
(0\pair{w,w},0\pair{w,w}) & 
          \mbox{if $(\exists w,x,y \in \sigmastar)\, (\exists m \in \nats)\, 
                  [a = 1\pair{w, 1x} \land z = 1\pair{y,0^m}]$} \\ 
(\epsilon , \epsilon) & \mbox{otherwise.}
\end{array}
\right.
\]
Note that $f \in \fp$.  Whenever there exists some string $b \in
\sigmastar$ for which $\sigma(a,b) = z$, it holds that
$\sigma(f(1,a,z)) = z$.  (If there is no such~$b$, it does not
matter what $f(1,a,z)$ outputs.)~~Hence $\sigma$ is not overstrong.~\qed

\bigskip

\noindent {\bf Acknowledgments:}
We thank Osamu Watanabe for
mentioning to us
the notions, different from those used here though
slightly reminiscent, from average-case theory, of
claw-free collections,
collision-free pseudorandom
generators, and
collision-free hash
functions.
We thank Chris Homan 
for suggesting 
overstrongness.

\end{document}